\documentclass[preprint,showkeys]{revtex4}

\usepackage{graphicx}

\begin{document}

\title{Effective potentials, thermodynamics, and
twisted mass quarks}
\author{Michael Creutz}
\affiliation{
Physics Department, Brookhaven National Laboratory\\
Upton, NY 11973, USA
}
\email{creutz@bnl.gov}

\begin{abstract}
Using effective Lagrangian arguments, I explore the qualitative
behavior expected at finite temperature for two flavor lattice QCD
formulated with Wilson fermions and a twisted mass term.  A rather
rich phase structure is predicted, exhibiting Aoki's parity violating
phase along with a deconfinement region forming a conical structure in
the space of coupling, hopping parameter, and twisted mass variables.
\end{abstract}


\maketitle

\section{Introduction}

Non-perturbative phenomena are crucial to understanding strong
interaction physics.  Two particularly powerful tools in this context
are lattice gauge theory and effective field theories.  Combining
these approaches can frequently give new insights.  For example,
effective Lagrangians can model the mechanisms for the lattice
artifacts that mutilate chiral symmetry, explaining
\cite{Creutz:1996bg,Sharpe:1998xm} such phenomena as the spontaneous
breaking of parity and charge conjugation discussed long ago by Aoki
\cite{Aoki:1983qi}.  Attempting to reduce artifacts stimulates the
development of new lattice actions, such as domain wall
\cite{Kaplan:1992bt,Furman:1994ky} or overlap
\cite{Narayanan:1994gw,Neuberger:1997fp} fermions.  The chiral
symmetry inherent in effective Lagrangian approaches also has been
crucial in exposing the issues inherent in the rooting procedure often
used to adjust the number of quark species \cite{Creutz:2007yg}.

One of the major successes of lattice QCD is the quantitative estimate
of the temperature for the transition from ordinary hadronic matter to
a plasma of quarks and gluons \cite{Heller:2006ub}.  Chiral symmetry
has an interesting interplay with this transition; indeed, it appears
that this symmetry is restored at the same temperature as
deconfinement.  Thus, finite temperature QCD is a natural playground
for gaining a better understanding of chiral symmetry.

This paper brings together the three topics of lattice artifacts,
chiral symmetry, and the deconfinement transition.  The presence of
lattice artifacts brings in new types of mass terms, one of which is
often referred to as a ``twisted'' mass \cite{Frezzotti:2003ni}.  The
main predictions are a rather intricate phase structure which can be
looked for in numerical simulations.  The phase structure with twisted
mass at zero temperature has been described in
Refs.~\cite{Sharpe:2004bv,Munster:2004am,Munster:2004wt}.  The main
addition here is the inclusion of the deconfinement transition in this
picture.

One might well ask why care about lattice artifacts?  Of course this
is necessary to understand the limitations of lattice simulations.  At
a more practical level, it is through understanding these effects that
one can explore the reasons for the failure of the rooting procedure
used with staggered quarks.  At the conceptual level, seeing the
distortions of chiral symmetry can help understand the nature of this
symmetry and how chiral anomalies work on the lattice.  These issues
are closely related to how quark masses are defined
\cite{Creutz:2003xc}.  One might also hope that understanding these
aspects can give insight into why gauge theories coupled to chiral
currents, such as in the standard model, are so hard to put on the
lattice.  And finally, we will see that the lattice artifacts give
rise to a rather amusing and somewhat complicated phase structure.

For simplicity, this discussion is restricted to QCD with two
degenerate quark flavors.  With three or more flavors the mass
``twisting'' is more complicated and not unique.  Furthermore, we only
consider the theory without any CP non-invariant term associated with
the gauge field topology.

Section \ref{wilson} starts off with an extremely brief review of the
Wilson fermion formulation \cite{Wilson:1975id}.  The discussion then
turns in section \ref{parameters} to the parameters of QCD,
emphasizing the non-linear mapping between the physical continuum
parameters and the bare parameters used on the lattice.  Then section
\ref{effective} reviews the lattice artifacts introduced through the
chiral symmetry breaking properties of the Wilson fermion action.
Section \ref{twisted} introduces the twisted mass term, which only has
meaning in the context of the lattice artifacts.  Section
\ref{deconfinement} brings finite temperature into the picture,
showing how deconfinement at high temperature interplays with the
previous lattice artifacts.  Section \ref{predictions} combines the
previous ideas to predict qualitatively the rich phase structure that
should appear in the space of lattice parameters.  In particular, the
deconfined phase should appear as an approximately conical structure
which should be looked for in simulations.  Finally remarks on some
open questions appear in section \ref{conclusions}.

\section{Wilson fermions}
\label{wilson}

\begin{figure*}
\centering
\includegraphics[width=3in]{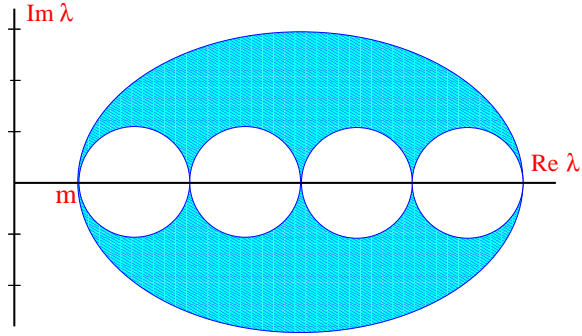}
\caption{ The eigenvalue spectrum of the free Wilson fermion operator
is a set of nested circles.  On turning on the gauge fields, some
eigenvalues drift into the open regions.  Some complex pairs can
collide and become real.  These are connected to gauge field topology.
}
\label{eigen1} 
\end{figure*}

For completeness this section gives a brief discussion of the Wilson
fermion approach.  The motivation is to overcome the doubling issue of
naive fermions by adding a momentum dependent mass to the extra
states.  Doubling arises because on the lattice physics becomes
naturally periodic in momenta, with momentum components being replaced
by trigonometric functions, i.e. $p_\mu \rightarrow {1\over 2a}
\sin(p_\mu a)$.  The problem is that this quantity is small not just
for small momentum, but also for $p_\mu\sim \pi/a$.  Adding a further
trigonometric behavior in the form of a momentum dependent mass, the
extraneous doublers can be made heavy.  For the free case, the
simplest version of Wilson fermions uses the Dirac operator in
momentum space
\begin{equation}
D_W(p)={1\over a}+{2K\over a}\sum_\mu \left[
i\gamma_\mu\sin(p_\mu a)-\cos(p_\mu a)\right].
\end{equation}
The physical fermion mass is read off from the small momentum behavior
as $m={1\over 2a}(1/K-8)$.  This vanishes at at $K=K_c=1/8$.  The eigenvalues of
this free operator lie on a set of ``nested circles,'' as sketched in
Fig.~\ref{eigen1}.

This discussion assumes that the gauge fields are formulated as usual
with group valued matrices on the lattice links.  To be specific one
can consider using the simple Wilson gauge action as a sum over
plaquettes
\begin{equation}
S_g={\beta\over 3}\sum_p {\rm Re\ Tr\ } U_p
\end{equation}
although this specific form is not essential the qualitative nature of
the phase diagram.  When the gauge fields are turned on, the dynamics
will move the fermion eigenvalues around, partially filling the holes
in eigenvalue pattern of Fig.~(\ref{eigen1}).  Some eigenvalues can
become real and are related to gauge field topology
\cite{Creutz:2006ts}.

Note that the basic lattice theory has two parameters parameters,
$\beta$ and $K$.  These are related to bare coupling, $\beta\sim
6/g_0^2$, and quark mass, $(1/K-1/K_c)\sim m_q$.  This will be
discussed further in the next section.

\section{Lattice versus continuum parameters}
\label{parameters}

The quark confining dynamics of the strong interactions, QCD, is a
remarkably economical theory in terms of the number of adjustable
parameters.  First of these is the overall strong interaction scale,
$\Lambda_{qcd}$.  This is scheme dependent, but once a renormalization
procedure has been selected, it is well defined.  It is not
independent of the coupling constant, the connection being fixed by
asymptotic freedom.  In addition, the theory can depend on the quark
mass, or more precisely the dimensionless ratio $m/\Lambda{qcd}$.  As
with the overall scale, the definition of $m$ is scheme dependent.
The two flavor theory with degenerate quarks has only one such mass
parameter.  On considering the theory at a finite physical
temperature, this adds another parameter, which can also be considered
in units of the overall scale, i.e. consider the ratio $T\over
\Lambda_{qcd}$.  Finally, to relate things to the theory with a
lattice cutoff, add a scale for this cutoff.  As with everything else,
measure this in units of the overall scale; so, the fourth parameter
is $a\Lambda_{qcd}$, where one can regard $a$ as the lattice spacing.

The goal here is to explore the lattice artifacts that arise with
Wilson fermions \cite{Wilson:1975id}.  On the lattice it is generally
easier to work directly with lattice parameters.  One of these is the
plaquette coupling $\beta$, which, with the usual conventions, is
related to the bare coupling $\beta=6/g_0^2$.  For the quarks, the
natural lattice quantity is the ``hopping parameter'' $K$.  To include
finite temperature effects, it is customary to work with a finite
temporal lattice of $N_t$ sites.  And finally, the connection with
physical scales appears via the lattice spacing $a$.

\begin{figure}
\centering
\includegraphics[width=3in]{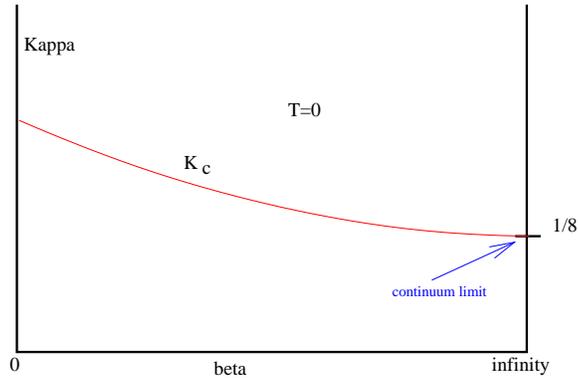}
\caption{ The continuum limit of lattice gauge theory with Wilson
fermions occurs at $\beta\rightarrow\infty$ and $K\rightarrow 1/8$.
Coming in from this point to finite beta is the curve $K_{c}(\beta)$,
representing the lowest phase transition in $K$ for fixed beta.  The
nature of this phase transition is a delicate matter, discussed in the
text.  }
\label{kbeta1}
\end{figure}

\begin{figure}
\centering
\includegraphics[width=3in]{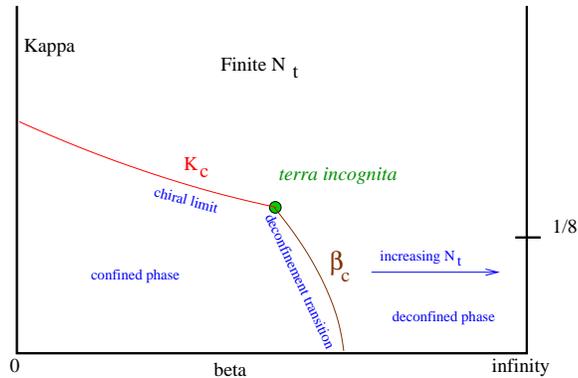}
\caption{ At finite $N_t$ the deconfinement phase transition appears
at a finite value for $\beta$ which depends on the hopping parameter
$K$.  For the continuum limit $N_t$ is taken to infinity and this
structure moves towards $\beta=\infty$.  The region above $K_c(\beta)$
is the object of later discussion.  }
\label{kbeta2}
\end{figure}

The set of physical parameters and the set of lattice parameters are,
of course, equivalent, and there is a well understood non-linear
mapping between them
\begin{equation}
\left\{{m\over\Lambda_{qcd}},{T\over\Lambda_{qcd}},
a\Lambda_{qcd}\right\}
 \longleftrightarrow \{\beta,K,N_t\}.
\end{equation}
Of course, to extract physical predictions we are interested in the
continuum limit $a\Lambda_{qcd}\rightarrow 0$.  For this, asymptotic
freedom tells us we must take $\beta\rightarrow \infty$ at a rate tied
to $\Lambda_{qcd}$.  Simultaneously we must take the hopping parameter
to a critical value.  With normal conventions, this takes
$K\rightarrow K_c\rightarrow 1/8$ at a rate tied to desired quark mass
$m$.  Finally, the number of temporal sites must also go to infinity
as $N_t= {1\over aT}$.  Fig.~\ref{kbeta1} sketches how the continuum
limit is taken in the $\beta,K$ plane for zero temperature.
Fig.~\ref{kbeta2} sketches how the basic phase diagram is modified for
a finite number of time slices.  Some early results on this structure
from numerical simulation appear in Ref.~\cite{Ukita:2006pc}.  The
subject of this paper is to further explore this phase diagram with
particular attention to hopping parameters larger than $K_c$.  This
discussion adds the possible twisted mass term (defined later) to the
exposition in \cite{Creutz:1996bg}.  Some very preliminary studies of
this system at finite temperature are presented in
Ref.~\cite{Ilgenfritz:2006tz}.

\section{Effective chiral Lagrangians and lattice artifacts}
\label{effective}

\begin{figure}
\centering
\includegraphics[width=3in]{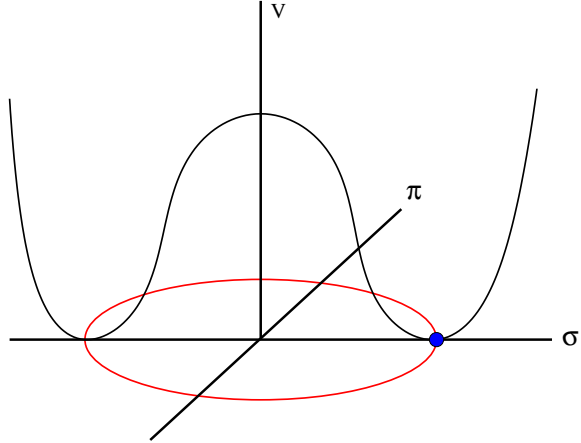}
\caption{
The symmetric ``Mexican hat'' potential for the pion and sigma fields
before mass terms and lattice artifacts are turned on.
}
\label{v0.eps}
\end{figure}

The use of effective field theories to describe the interactions of
the pseudo-scalar mesons is an old and venerable topic.  Here we will
only discuss the simplest form for the two flavor theory, adding terms
that mimic the expected lattice artifacts.  The language is framed in
terms of the isovector pion field $\vec\pi\sim i\overline\psi \gamma_5
\vec \tau \psi$ and the scalar sigma $\sigma\sim \overline\psi \psi$.
the starting point for this discussion is the canonical ``Mexican
hat'' potential
\begin{equation}
V_0=\lambda (\sigma^2+\vec\pi^2-v^2)^2
\end{equation}
schematically sketched in Fig.~\ref{v0.eps}.  The potential has a
symmetry under $O(4)$ rotations amongst the pion and sigma fields
expressed as the four vector $\Sigma=(\sigma,\vec\pi)$.  This
represents the axial symmetry of the underlying quark theory.

As usually considered, this theory is taken with the minimum for the
potential occurring at a non-vanishing value for the fields.  This is
a classic example of spontaneous symmetry breaking, and the vacuum is
conventionally selected to lie in the sigma direction with
$\langle\sigma\rangle > 0$.  The pions are then Goldstone bosons of
the theory, being massless because the potential provides no barrier
to oscillations of the fields in the pionic directions.

\begin{figure}
\centering
\includegraphics[width=3in]{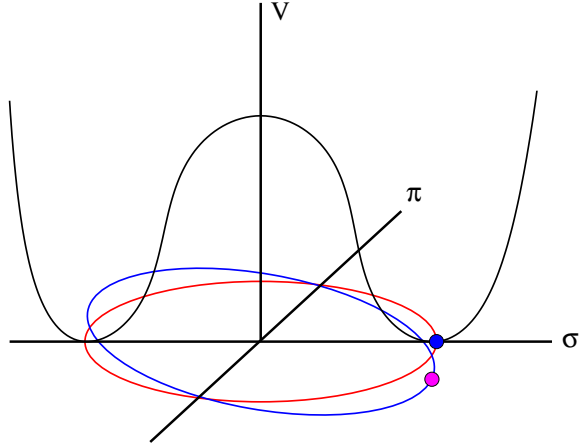}
\caption{
A mass terms acts by tilting the
potential, thus selecting a unique vacuum.
}
\label{tilt}
\end{figure}

With this potential, it is natural to include a quark mass by adding a
constant times the sigma field
\begin{equation}
 V_1=-m\sigma
\end{equation}
This explicitly breaks the chiral symmetry by ``tilting'' the
potential as sketched in Fig.~\ref{tilt}.  This selects a unique
vacuum which, for $m>0$, gives a positive expectation for sigma.  In
the process the pions gain a mass, with $m_\pi^2\sim m$.

There is an obvious freedom in the selection of the mass term.
Because of the symmetry of $V_0$, it does not physically matter in
which direction we tilt the vacuum.  In particular, a mass term of
form
\begin{equation}
m\sigma\rightarrow m\cos(\theta) \sigma + m\sin(\theta)\pi_3
\label{mrot}
\end{equation}
should give equivalent physics for any $\theta$.  However, as we will
see, lattice artifacts can break this symmetry, introducing the
possibility of physics at finite lattice spacing which
depends on this angle.  The second term in this equation is what we
will call the ``twisted mass.''

The Wilson term inherently breaks chiral symmetry.  This will give
rise to various modifications of the effective potential.  The first
correction is expected to be an additive contribution to the quark
mass, i.e. an additional tilt to the potential.  This means that the
critical kappa, defined as the smallest kappa where a singularity is
found in the $\beta,K$ plane, will move away from the limiting value
of 1/8.  Thus we introduce the function $K_c(\beta)$ and imagine that
the mass term is modeled with the form
\begin{equation}
m\rightarrow c_1 (1/K-1/K_c(\beta))
\label{c1}
\end{equation}

In general the modification of the effective potential will have
higher order corrections.  A natural way to include such is as an
expansion in the chiral fields.  With this motivation we include a
term in the potential of form
\begin{equation}
c_2 \sigma^2
\end{equation}
Such a term was considered in \cite{Creutz:1996bg,Sharpe:1998xm}.  The
predicted phase structure depends qualitatively on on the sign of
$c_2$, but a priori we have no information on this. Indeed, as it is a
lattice artifact, it is expected that this sign might depend on the
choice of gauge action.  Note that we could have added a term like
$\vec\pi^2$, but this is essentially equivalent since
$\vec\pi^2=(\sigma^2+\vec\pi^2)-\sigma^2$, and the first term here can
be absorbed, up to an irrelevant constant, into the starting Mexican
hat potential.

\begin{figure}
\centering
\includegraphics[width=3in]{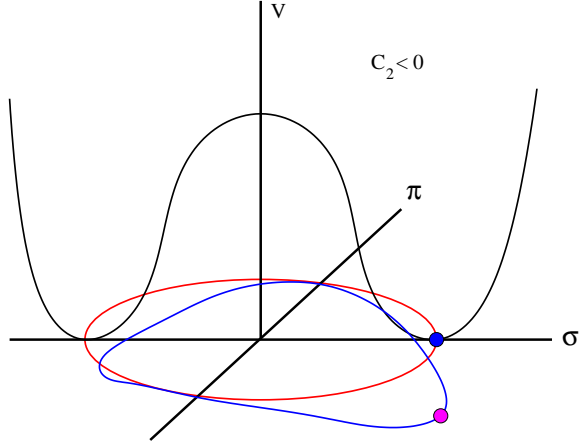}
\caption{
Lattice artifacts could quadratically warp the effective potential.
If this warping is downward in the sigma direction, the chiral
transition becomes first order without the pions becoming massless.
}
\label{c2lt0}
\end{figure}

First consider the case when $c_2$ is less than zero, thus lowering
the potential energy when the field points in the positive or negative
sigma direction.  This quadratic warping helps to stabilize the sigma
direction, as sketched in Fig.~\ref{c2lt0}, and the pions cease to be
true Goldstone bosons when the quark mass vanishes.  Instead, as the
mass passes through zero, we have a first order transition as the
expectation of $\sigma$ jumps from positive to negative.  This jump
occurs without any physical particles becoming massless.

\begin{figure}
\centering
\includegraphics[width=3in]{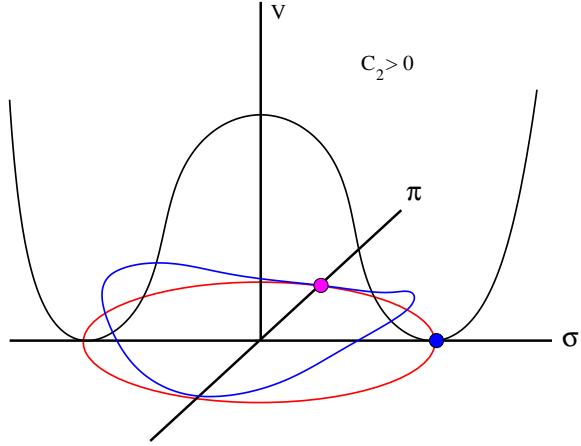}
\caption{
If the lattice artifacts warping the potential upward in the sigma
direction, the chiral transition splits into two second order
transitions separated by a phase where the pion field has an
expectation value.
}
\label{c2gt0}
\end{figure}

Things get a bit more complicated if $c_2>0$, as sketched in
Fig.~\ref{c2gt0}.  In that case the chiral transition splits into two
second order transitions separated by phase with an expectation for
the pion field, i.e. $\langle \vec\pi \rangle \ne 0$.  Since the pion
field has odd parity and charge conjugation as well as carries
isospin, all of these symmetries are spontaneously broken in the
intermediate phase.  As isospin is a continuous group, this phase will
exhibit Goldstone bosons.  The number of these is two, representing
the two flavor generators orthogonal to the direction of the
expectation value.  If higher order terms do not change the order of
the transitions, there will be a third massless particle exactly at
the transition endpoints.  In this way the theory does acquire three
massless pions exactly at the transitions, as discussed by Aoki
\cite{Aoki:1983qi}.  The intermediate phase is usually referred to as
the ``Aoki phase.''

\begin{figure}
\centering
\includegraphics[width=3in]{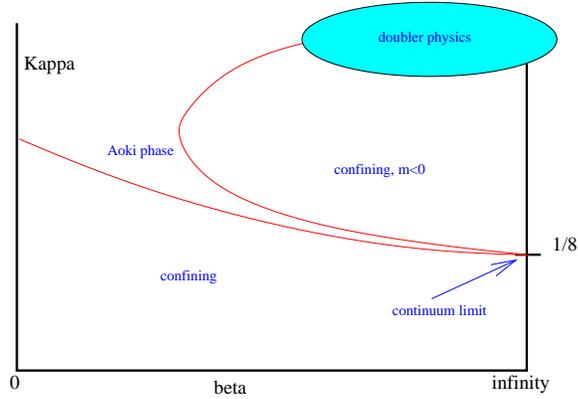}
\caption{The qualitative structure of the $\beta,K$ plane including
the possibility of an Aoki phase.  }
\label{kbeta3}
\end{figure}

To include these ideas in the effective model, add the $c_2$ term to
the potential
\begin{equation}
V(\vec\pi,\sigma,L)=\lambda(\sigma^2+\vec\pi^2-v^2)^2-c_1(1/K-1/K_c(\beta))\sigma
+c_2\sigma^2.
\end{equation}
Assuming the $c_2>0$ case, Fig.~\ref{kbeta3} shows the qualitative
phase diagram expected.

\section{Twisted mass}
\label{twisted}

The $c_2$ term breaks the equivalence of different chiral directions.
This means that physics will indeed depend on the angle $\theta$ if
one takes a mass term of the form in Eq.~(\ref{mrot}).
Consider complexifying the fermion mass in the usual way
\begin{equation}
m\overline\psi\psi\rightarrow{1\over 2}(m\overline\psi_L\psi_R
+m^*\overline\psi_R\psi_L)
\end{equation}
The rotation of  Eq.~(\ref{mrot}) is equivalent to giving the up and
down quark masses opposite phases
\begin{eqnarray}
m_u\rightarrow e^{+i\theta}m_u\\ 
m_d\rightarrow e^{-i\theta}m_d
\end{eqnarray}
Thus motivated, we can consider adding a new mass term to the lattice theory
\begin{equation}
\mu\ i\overline\psi \tau_3\gamma_5\psi \sim \mu\pi_3
\end{equation}
This extends our effective potential to
\begin{equation}
V(\vec\pi,\sigma)=\lambda(\sigma^2+\vec\pi^2-v^2)^2-c_1(1/K-1/K_c(\beta))\sigma
+c_2\sigma^2-\mu \pi_3
\end{equation}
The twisted mass represents the addition of a ``magnetic field''
conjugate to the order parameter for the Aoki phase.

\begin{figure}
\centering
\includegraphics[width=3in]{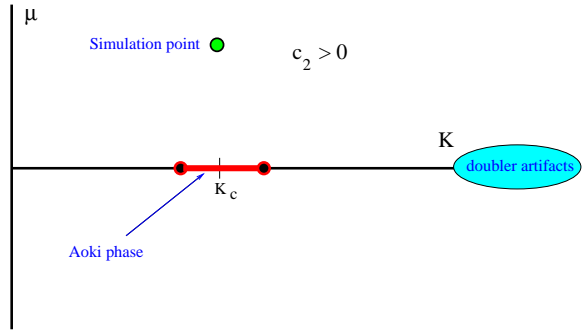}
\caption{
Continuing around the Aoki phase with twisted mass.  This sketch
considers the case $c_2>2$ where the parity broken phase extends over
a region along the kappa axis.
}
\label{tm1}
\end{figure}

\begin{figure}
\centering
\includegraphics[width=3in]{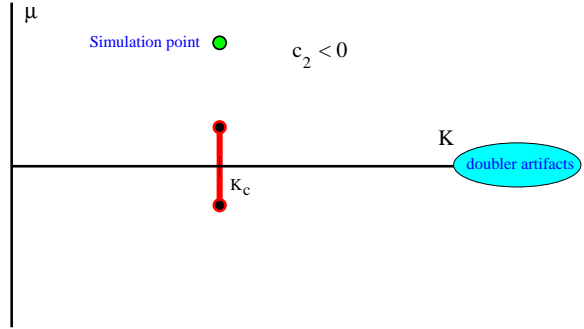}
\caption{
As in Fig.~\ref{tm1}, but for the case $c_2<0$ so the chiral
transition on the kappa axis becomes first order.
}
\label{tm2}
\end{figure}

There are a variety of motivations for adding such a term to our
lattice action \cite{Frezzotti:2003ni,Munster:2004wt}.  Primary among
them is that $O(a)$ lattice artifacts can be arranged to cancel.  With
two flavors of conventional Wilson fermions, these effects change sign
on going from positive to negative mass, and if we put all the mass
into the twisted term we are half way between.  It should be noted
that this cancellation only occurs when all the mass comes from the
twisted term; for other combinations with a traditional mass term,
some lattice artifacts of $O(a)$ will survive.  Also, although it
looks like we are putting phases into the quark masses, these cancel
between the two flavors, and the resulting fermion determinant remains
positive.  Furthermore, the algorithm is considerably simpler and
faster than either overlap \cite{Narayanan:1994gw,Neuberger:1997fp} or
domain wall \cite{Kaplan:1992bt,Furman:1994ky} fermions while avoiding
the diseases of staggered quarks \cite{Creutz:2007yg}.  Another nice
feature of adding a twisted mass term is that it allows a better
understanding of the Aoki phase and shows how to continue around it.
Figs.~\ref{tm1} and \ref{tm2} show how this works for the case $c_2>0$
and $c_2<0$, respectively.

Of course some difficulties come along with these advantages.  First,
one needs to know $K_c$.  Indeed, with the Aoki phase present, the
definition of this quantity is not unique.  Second, it is not clear
how to extend twisting to odd numbers of flavors, where the
$m\leftrightarrow -m$ symmetry is broken by anomalies; the negative
mass case then corresponds to the strong CP angle $\Theta$ being
$\pi$, where one expects a spontaneous CP violation surviving in the
continuum limit.  And finally, the mass needs to be larger than the
$c_2$ artifacts.  Indeed, as Figs.~\ref{tm1} and \ref{tm2} suggest, if
it is not, then one is really studying the physics of the Aoki phase,
not the correct continuum limit.  This also has implications for how
close to the continuum one must be to study this structure; in
particular, one must have $\beta$ large enough so the Aoki phase does
not extend into the doubler region.

\section{Deconfinement at finite temperature}
\label{deconfinement}

Now we bring finite temperature and the deconfinement transition into
the discussion.  To treat temperature in the path integral formalism,
one keeps the temporal extent of the system finite.  The temporal
boundary conditions should be periodic for bosonic fields and
anti-periodic for fermionic ones.  The usual order parameter for the
transition is the Wilson-Polyakov line.  In continuum language this
involves an integral that loops around the temporal direction
\begin{equation}
L(\vec x)={\rm Re\ Tr} {\cal T} \exp(i\int_0^{1/T} A_0(\vec x) dt).
\end{equation}
where $\cal T$ denotes time ordering.  Of course this quantity needs
renormalization for definition.  On the lattice this order parameter
reduces to the trace of an ordered product of link variables that
wraps similarly around the system
\begin{equation}
L(\vec x)={\rm Re\ Tr} \prod_t U_{t+1,t}(\vec x)
\end{equation}
where the product is again time ordered.  Physically, one can regard
the expectation of $L$ as the exponentiated energy of a stationary
source with quark quantum numbers.  When $\langle L\rangle$ vanishes,
such a source is confined.

The pure gauge theory, where quarks are left out, has a symmetry under
taking $L\rightarrow e^{2\pi i/3} L$.  This can be seen since taking
all time-like links at a single fixed time slice and multiplying them
by a third root of unity (which is, of course, an $SU(3)$ element)
leaves the action invariant.  At low temperatures, when confinement is
manifest, the vanishing of $\langle L\rangle$ corresponds to the
symmetry being unbroken.  On the other hand, when the temperature is
high, it is expected that the symmetry will be spontaneously broken,
corresponding to a deconfined phase.  When quarks are introduced with
a finite mass, the symmetry ceases to be exact.  Physically, dynamical
quarks can screen a fixed source, allowing it to have finite energy
even if confinement persists.  Because of this breaking, the
deconfinement transition need not be a true phase transition, and
could be only crossover, i.e. a rapid change of behavior over a small
but finite range of temperature.

An alternative order parameter for the deconfinement transition is the
quark condensate $\langle\overline\psi\psi\rangle\sim
\langle\sigma\rangle$.  At vanishing quark mass, this marker of chiral
symmetry breaking is expected to vanish at high temperature,
indicative of the evolution of the effective potential
$V(\vec\pi,\sigma)$ to a state with a single minimum.

Numerous numerical simulations \cite{Heller:2006ub} have clearly
demonstrated that both $\langle L \rangle$ and
$\langle\overline\psi\psi\rangle$ do show a rapid change over a single
small region of temperature.  For large quark mass this transition
becomes first order, much as seen in the three state Potts model,
which has the same $Z_3$ symmetry.  At $m=0$ it is generally expected
that the transition becomes second order and in the same universality
class as the $O(4)$ sigma model.  However this is not proven and there
are some hints \cite{D'Elia:2007fh} that the transition may be first
order.  Between $m=0$ and $m=\infty$ it appears that for a large
region the transition reduces to a cross-over.

Various effective models have been invoked to mimic this behavior
\cite{Svetitsky:1982gs,Gocksch:1991up,Ogilvie:1983ss,Pisarski:1983db}.
These are based on treating the Wilson line as an effective complex
field $L(\vec x)$ in the three spatial dimensions.  A simple starting
potential is
\begin{equation}
V(L)=\alpha_1|L|^4+\alpha_2(T_c-T) |L|^2 
-\alpha_3{\rm Re} (L^3) -\alpha_4{\rm Re} L
\end{equation}
Here the $\alpha_1>0$ term serves to keep the system stable.  The
$\alpha_2>0$ term is the basic driving term for the transition; when
the temperature is below $T_c$ the potential has a unique minimum
corresponding to a small or vanishing value for the expectation of the
effective field.  But when the temperature exceeds $T_c$, this minimum
inverts and we can have a spontaneous symmetry breaking with $\langle
L\rangle$ developing a large value.  The $\alpha_3$ term serves to
reduce the symmetry to the physical $Z_3$ rather than the $U(1)$
present with the first two terms alone.  This also drives the
transition towards first order because of its cubic nature.  Finally,
the $\alpha_4$ term is present to represent the fact that dynamical
quarks break the symmetry since they can screen other fundamental
charges.  This term can also soften the transition to a crossover
since it allows the order parameter to have a small expectation in the
low temperature region.  To proceed, rewrite this potential as a
function of the lattice parameters, i.e.
\begin{equation}
T_c-T \rightarrow \beta_c(K,N_t)-\beta;
\end{equation} 
where $\beta_c(K,N_t)$ is the value of $\beta$ where deconfinement
sets in at a fixed $N_t$ and hopping parameter.  Thus we have
\begin{equation}
V(L)=\alpha_1|L|^4+\alpha_2(\beta_c(K,N_t)-\beta) |L|^2 
-\alpha_3{\rm Re} (L^3) -\alpha_4{\rm Re} L.
\end{equation}

With effective models for both the chiral fields and the Wilson line,
it is natural to unite them into a single effective model involving
all fields \cite{Mocsy:2003qw,Ratti:2005jh}.  Combining the previous
potentials, consider
\begin{equation}
\matrix{
V(\vec\pi,\sigma,L)&=\lambda(\sigma^2+\vec\pi^2-v^2)^2-c_1(1/K-1/K_c(\beta))
\sigma
+c_2\sigma^2-\mu \pi_3\cr
&+\alpha_1|L|^4+\alpha_2(\beta_c(K,N_t)-\beta) |L|^2 
-\alpha_3{\rm Re} (L^3) -\alpha_4{\rm Re} L\cr
&+\alpha_5 |L|^2(\sigma^2+\vec\pi^2)\cr
}
\end{equation}
Here the $\alpha_5$ term serves to couple the chiral fields with the
loop field.  Through this term, a jump in $|L|$ can turn off the
chiral symmetry breaking.  The potential involves two unknown
functions: $K_c(\beta)$ and $\beta_c(K,N_t)$.  For simplicity in the
following discussion, ignore any possible $K_c$ dependence on $N_t$
before deconfinement takes place.

\section{Predictions}
\label{predictions}

\begin{figure}
\centering
\includegraphics[width=2.5in]{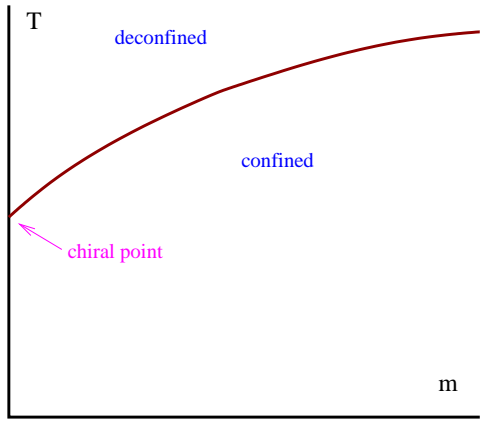}
\caption{
The deconfinement transition increases monotonically with the quark
mass.
}
\label{tcm}
\end{figure}

\begin{figure}
\centering
\includegraphics[width=3in]{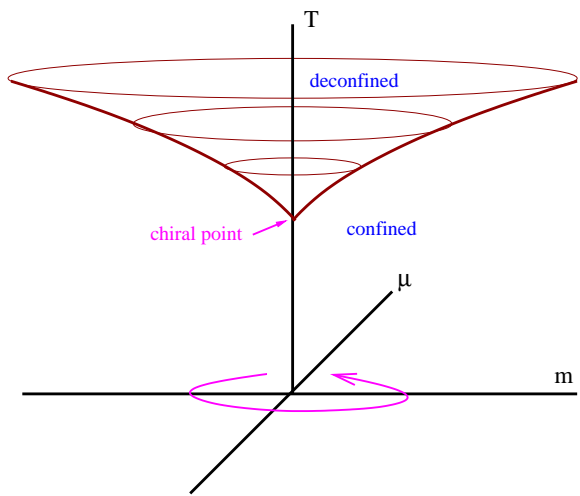}
\caption{
Rotating the transition line of Fig.~\ref{tcm} into twisted mass space
gives an ``inverted umbrella'' or ``conical'' structure.
}
\label{tcm2}
\end{figure}

This combined effective potential allows us to sketch qualitatively
much of the expected phase diagram for twisted mass Wilson fermions at
finite temperature.  To understand the structure it is useful to first
extract some qualitative features of the continuum physics using the
numerically supported fact that increasing the quark mass increases
the deconfinement temperature, as sketched schematically in
Fig.~\ref{tcm}.  Since continuum physics is independent of twisting,
if we introduce a twisted mass term we obtain an ``inverted umbrella''
or ``conical'' structure in $m,\mu,T$ space.  This structure is
symmetric under rotations about the $T$ axis, as sketched in
Fig.~\ref{tcm2}.  Of course, as mentioned earlier, the transition need
not be a true singularity, but could, for some mass range, just be a
rapid crossover.

\begin{figure}
\centering
\includegraphics[width=3in]{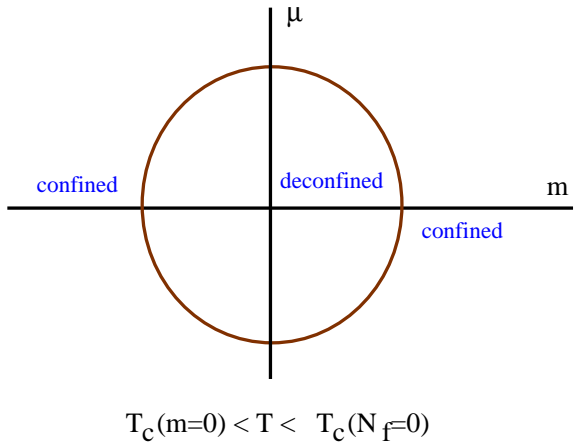}
\caption{In the continuum the deconfinement transition is a simple
circle in $m,\mu$ space.  }
\label{tm2.5}
\end{figure}

Now consider an intermediate temperature where the massless theory is
deconfined but at some finite mass the theory is still confined.  In
particular, consider a horizontal slice intersecting the ``cone'' of
Fig.~\ref{tcm2}.  At this temperature the phase diagram in the $m,\mu$
plane is simply a circle as sketched in Fig.~\ref{tm2.5}; again, this
is an immediate consequence of the continuum physics being independent
of twisting.  Note that this figure makes it clear that for this two
flavor theory physics at some mass $m$ is equivalent to physics at
$-m$, and the two regions are connected by rotations in the $m,\mu$
plane.

\begin{figure}
\centering
\includegraphics[width=3in]{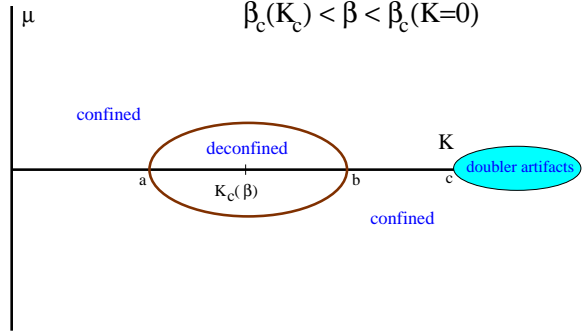}
\caption{
At fixed $N_t$ and for a value of $\beta$ where the infinite mass
theory is confined but the massless theory is in the deconfined phase,
we expect the deconfined phase to assume an approximately elliptical
structure in the $K\mu$ plane.
}
\label{tm3}
\end{figure}

On the lattice we should expect a similar structure, at least near the
continuum limit.  However, twisting the mass is no longer an exact
symmetry; so, we can expect the circles of equivalent physics to be
distorted.  Also, in the deconfined region we expect the effective
potential to have a single minimum, so that the Aoki phase will wash
out after deconfinement.  Actually, this disappearance of the Aoki
structure at high temperature should be independent of lattice issues
such as the sign of $c_2$.  Qualitatively this means that at some
fixed $\beta$ and $N_t$ one should see a structure similar to that
sketched in Fig.~\ref{tm3}.  As mentioned earlier, $\beta$ will need
to be large enough that the Aoki phase structure is well separated
from the doubler region.  As one comes closer to the continuum limit,
this structure should become increasingly precisely an ellipse with
constant $\sqrt {\mu^2+ c_1^2 (1/K-1/K_c)^2}$ with $c_1$ from
Eq.~\ref{c1}.  Presumably this can be verified in simulations for
small $\mu$ near the lower deconfinement transition.  Working near the
upper transition may be more difficult in practice due to small
eigenvalues of the Dirac operator for supercritical kappa.

\begin{figure}
\centering
\includegraphics[width=3in]{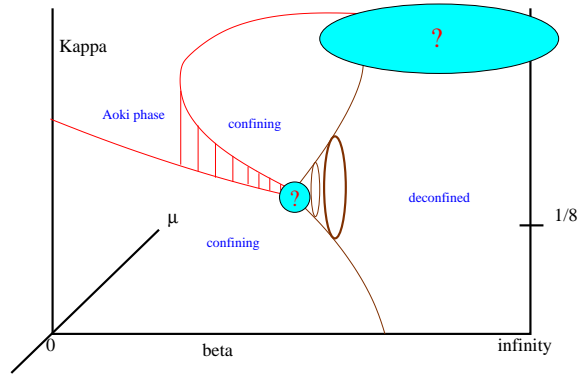}
\caption{ The predicted phase diagram in $\beta,K,\mu$ space for fixed
$N_t$.  Note how one can smoothly continue from the conventional
confining phase to another confining phase in the supercritical
hopping region.  }
\label{kbeta4}
\end{figure}

Extending this picture to the full $\beta,K,\mu$ space at fixed $N_t$,
we expect the Aoki phase to dissolve into a conical structure as
sketched in Fig.~\ref{kbeta4}.  Just how the deconfining cone joins on
to the end of the Aoki phase presumably is rather sensitive to
dynamical details.

\section{Conclusions and open questions}
\label{conclusions}

We have seen that Wilson fermions at finite temperature and with a
twisted mass term can give rise to a fascinatingly complex phase
structure, much of it due to lattice artifacts.  This suggests that it
could be interesting to study the supercritical hopping region in more
detail.  In particular a second deconfinement line is expected, as
least close to the continuum limit.  The twisted mass addition allows
us to smoothly connect two deconfined phases.  This connection should
expose an approximately elliptical structure in the $K,\mu$ plane at
fixed $\beta$.

This picture raises some interesting questions.  One concerns the
three flavor theory.  In this case the parity broken phase becomes
physical.  Indeed, three degenerate quarks of negative mass represent
QCD with a strong CP angle $\theta=\pi$, for which spontaneous
breaking of CP is expected.  With three flavors the twisting process
is not unique, with possible twists in the $\lambda_3$ or $\lambda_8$
directions.  For example, using only $\lambda_3$ would suggest a
twisted mass of form $m_u\sim e^{2\pi i/3}$, $m_d\sim e^{-2\pi i/3}$,
$m_s\sim 1$.

Another interesting case is one flavor QCD \cite{Creutz:2006ts}.  In
this situation the anomaly removes all chiral symmetry, and the quark
condensate loses meaning as an order parameter.  The critical value of
kappa where the mass gap disappears is decoupled from the point of
zero physical quark mass.  There is a parity broken phase, but it
occurs only at sufficiently negative mass.  And from the point of view
of twisting the mass, without chiral symmetry there is nothing to
twist.

\section*{Acknowledgments}
This manuscript has been authored under contract number
DE-AC02-98CH10886 with the U.S.~Department of Energy.  Accordingly,
the U.S. Government retains a non-exclusive, royalty-free license to
publish or reproduce the published form of this contribution, or allow
others to do so, for U.S.~Government purposes.


\begin{thebibliography}{99}

%\cite{Creutz:1996bg}
\bibitem{Creutz:1996bg}
  M.~Creutz,
  %``Wilson fermions at finite temperature,''
  arXiv:hep-lat/9608024.
  %%CITATION = HEP-LAT/9608024;%%

%\cite{Sharpe:1998xm}
\bibitem{Sharpe:1998xm}
  S.~R.~Sharpe and R.~L.~Singleton,
  %``Spontaneous flavor and parity breaking with Wilson fermions,''
  Phys.\ Rev.\  D {\bf 58}, 074501 (1998)
  [arXiv:hep-lat/9804028].
  %%CITATION = PHRVA,D58,074501;%%

%\cite{Aoki:1983qi}
\bibitem{Aoki:1983qi}
  S.~Aoki,
  %``New Phase Structure For Lattice QCD With Wilson Fermions,''
  Phys.\ Rev.\  D {\bf 30}, 2653 (1984).
  %%CITATION = PHRVA,D30,2653;%%

%\cite{Kaplan:1992bt}
\bibitem{Kaplan:1992bt}
  D.~B.~Kaplan,
  %``A Method for simulating chiral fermions on the lattice,''
  Phys.\ Lett.\  B {\bf 288}, 342 (1992)
  [arXiv:hep-lat/9206013].
  %%CITATION = PHLTA,B288,342;%%

%\cite{Furman:1994ky}
\bibitem{Furman:1994ky}
  V.~Furman and Y.~Shamir,
  %``Axial Symmetries In Lattice QCD With Kaplan Fermions,''
  Nucl.\ Phys.\  B {\bf 439}, 54 (1995)
  [arXiv:hep-lat/9405004].
  %%CITATION = NUPHA,B439,54;%%

%\cite{Narayanan:1994gw}
\bibitem{Narayanan:1994gw}
  R.~Narayanan and H.~Neuberger,
  %``A Construction of lattice chiral gauge theories,''
  Nucl.\ Phys.\  B {\bf 443}, 305 (1995)
  [arXiv:hep-th/9411108].
  %%CITATION = NUPHA,B443,305;%%

%\cite{Neuberger:1997fp}
\bibitem{Neuberger:1997fp}
  H.~Neuberger,
  %``Exactly massless quarks on the lattice,''
  Phys.\ Lett.\  B {\bf 417}, 141 (1998)
  [arXiv:hep-lat/9707022].
  %%CITATION = PHLTA,B417,141;%%

%\cite{Creutz:2007yg}
\bibitem{Creutz:2007yg}
  M.~Creutz,
  %``The evil that is rooting,''
  Phys.\ Lett.\  B {\bf 649}, 230 (2007)
  [arXiv:hep-lat/0701018].
  %%CITATION = PHLTA,B649,230;%%

%\cite{Heller:2006ub}
\bibitem{Heller:2006ub}
  U.~M.~Heller,
  %``Recent progress in finite temperature lattice QCD,''
  PoS {\bf LAT2006}, 011 (2006)
  [arXiv:hep-lat/0610114].
  %%CITATION = POSCI,LAT2006,011;%%

%\cite{Frezzotti:2003ni}
\bibitem{Frezzotti:2003ni}
  R.~Frezzotti and G.~C.~Rossi,
  %``Chirally improving Wilson fermions. I: O(a) improvement,''
  JHEP {\bf 0408}, 007 (2004)
  [arXiv:hep-lat/0306014];
  %%CITATION = JHEPA,0408,007;%%
%\cite{Frezzotti:2004wz}
%\bibitem{Frezzotti:2004wz}
  R.~Frezzotti and G.~C.~Rossi,
  %``Chirally improving Wilson fermions. II: Four-quark operators,''
  JHEP {\bf 0410}, 070 (2004)
  [arXiv:hep-lat/0407002];
  %%CITATION = JHEPA,0410,070;%%
%\cite{Frezzotti:2005zm}
%\bibitem{Frezzotti:2005zm}
  R.~Frezzotti and G.~Rossi,
  %``Chirally improving Wilson fermions. III: The Schroedinger functional,''
  arXiv:hep-lat/0507030;
  %%CITATION = HEP-LAT/0507030;%%
%\cite{Farchioni:2005bh}
%\bibitem{Farchioni:2005bh}
  F.~Farchioni {\it et al.},
  %``Numerical simulations with two flavours of twisted-mass Wilson quarks  and
  %DBW2 gauge action,''
  Eur.\ Phys.\ J.\  C {\bf 47}, 453 (2006)
  [arXiv:hep-lat/0512017].
  %%CITATION = EPHJA,C47,453;%%
%\cite{Boucaud:2007uk}
%\bibitem{Boucaud:2007uk}
  Ph.~Boucaud {\it et al.}  [ETM Collaboration],
  %``Dynamical twisted mass fermions with light quarks,''
  arXiv:hep-lat/0701012.
  %%CITATION = HEP-LAT/0701012;%%

%\cite{Sharpe:2004bv}
\bibitem{Sharpe:2004bv}
  S.~R.~Sharpe and J.~M.~S.~Wu,
  %``Applying chiral perturbation to twisted mass lattice QCD,''
  Nucl.\ Phys.\ Proc.\ Suppl.\  {\bf 140}, 323 (2005)
  [arXiv:hep-lat/0407035].
  %%CITATION = NUPHZ,140,323;%%

%\cite{Munster:2004am}
\bibitem{Munster:2004am}
  G.~Munster,
  %``On the phase structure of twisted mass lattice QCD,''
  JHEP {\bf 0409}, 035 (2004)
  [arXiv:hep-lat/0407006].
  %%CITATION = JHEPA,0409,035;%%

%\cite{Munster:2004wt}
\bibitem{Munster:2004wt}
  G.~Munster, C.~Schmidt and E.~E.~Scholz,
  %``Chiral perturbation theory for twisted mass QCD,''
  Nucl.\ Phys.\ Proc.\ Suppl.\  {\bf 140}, 320 (2005)
  [arXiv:hep-lat/0409066].
  %%CITATION = NUPHZ,140,320;%%

%\cite{Creutz:2003xc}
\bibitem{Creutz:2003xc}
  M.~Creutz,
  %``Ambiguities in the up-quark mass,''
  Phys.\ Rev.\ Lett.\  {\bf 92}, 162003 (2004)
  [arXiv:hep-ph/0312225].
  %%CITATION = PRLTA,92,162003;%%

%\cite{Wilson:1975id}
\bibitem{Wilson:1975id}
  K.~G.~Wilson,
  %``Quarks And Strings On A Lattice,''
%\href{http://www.slac.stanford.edu/spires/find/hep/www?r=clns-321}
%{SPIRES entry}
in {\it New Phenomena In Subnuclear Physics. Part A. Proceedings of the
First Half of the 1975 International School of Subnuclear Physics,}
Erice, Sicily, July 11 - August 1, 1975, ed. A.~Zichichi, Plenum
Press, New York, 1977, p.~69.

%\cite{Creutz:2006ts}
\bibitem{Creutz:2006ts}
  M.~Creutz,
  %``One flavor QCD,''
  arXiv:hep-th/0609187.
  %%CITATION = HEP-TH/0609187;%%

%\cite{Ukita:2006pc}
\bibitem{Ukita:2006pc}
  N.~Ukita, S.~Ejiri, T.~Hatsuda, N.~Ishii, Y.~Maezawa, S.~Aoki and K.~Kanaya,
  %``Finite temperature phase transition of two-flavor QCD with an improved
  %Wilson quark action,''
  PoS {\bf LAT2006}, 150 (2006)
  [arXiv:hep-lat/0610038].
  %%CITATION = POSCI,LAT2006,150;%%

%\cite{Ilgenfritz:2006tz}
\bibitem{Ilgenfritz:2006tz}
  E.~M.~Ilgenfritz, M.~Muller-Preussker, A.~Sternbeck, K.~Jansen, I.~Wetzorke, M.~P.~Lombardo and O.~Philipsen,
  %``Twisted mass QCD thermodynamics: First results on apeNEXT,''
  arXiv:hep-lat/0610112.
  %%CITATION = HEP-LAT/0610112;%%



%\cite{D'Elia:2007fh}
\bibitem{D'Elia:2007fh}
  M.~D'Elia, A.~Di Giacomo and C.~Pica,
  %``The thermodynamics of N(f) = 2 QCD,''
  Nucl.\ Phys.\ Proc.\ Suppl.\  {\bf 164}, 248 (2007).
  %%CITATION = NUPHZ,164,248;%%

%\cite{Svetitsky:1982gs}
\bibitem{Svetitsky:1982gs}
  B.~Svetitsky and L.~G.~Yaffe,
  %``Critical Behavior At Finite Temperature Confinement Transitions,''
  Nucl.\ Phys.\  B {\bf 210}, 423 (1982).
  %%CITATION = NUPHA,B210,423;%%
%\cite{Gocksch:1991up}

\bibitem{Gocksch:1991up}
  A.~Gocksch,
  %``Chiral symmetry in hot QCD,''
  Phys.\ Rev.\ Lett.\  {\bf 67}, 1701 (1991).
  %%CITATION = PRLTA,67,1701;%%

%\cite{Ogilvie:1983ss}
\bibitem{Ogilvie:1983ss}
  M.~Ogilvie,
  %``An Effective Spin Model For Finite Temperature QCD,''
  Phys.\ Rev.\ Lett.\  {\bf 52}, 1369 (1984).
  %%CITATION = PRLTA,52,1369;%%

%\cite{Pisarski:1983db}
\bibitem{Pisarski:1983db}
  R.~D.~Pisarski,
  %``Finite Temperature QCD At Large N,''
  Phys.\ Rev.\  D {\bf 29}, 1222 (1984).
  %%CITATION = PHRVA,D29,1222;%%

%\cite{Mocsy:2003qw}
\bibitem{Mocsy:2003qw}
  A.~Mocsy, F.~Sannino and K.~Tuominen,
  %``Confinement versus chiral symmetry,''
  Phys.\ Rev.\ Lett.\  {\bf 92}, 182302 (2004)
  [arXiv:hep-ph/0308135].
  %%CITATION = PRLTA,92,182302;%%

%\cite{Ratti:2005jh}
\bibitem{Ratti:2005jh}
  C.~Ratti, M.~A.~Thaler and W.~Weise,
  %``Phases of QCD: Lattice thermodynamics and a field theoretical model,''
  Phys.\ Rev.\  D {\bf 73}, 014019 (2006)
  [arXiv:hep-ph/0506234].
  %%CITATION = PHRVA,D73,014019;%%

\end{thebibliography}
\end{document}